# Title
High-Density EEG Enables the Fastest Visual Brain-Computer Interfaces

# Authors
Gege Ming[1], Weihua Pei[2,3], Sen Tian[4], Xiaogang Chen[5], Xiaorong Gao[1*], Yijun Wang[2,3,6*]

# Affiliations
[1]Department of Biomedical Engineering, Tsinghua University, Beijing 100084, China.
[2]Laboratory of Solid-State Optoelectronics Information Technology, Institute of Semiconductors, Chinese Academy of Sciences, Beijing 100083, China.
[3]School of Future Technology, University of Chinese Academy of Sciences, Beijing 100049, China.
[4]Suzhou Nianji Intelligent Technology Co., Ltd., Suzhou 215133, China.
[5]Institute of Biomedical Engineering, Chinese Academy of Medical Sciences and Peking Union Medical College, Tianjin 300192, China.
[6]Chinese Institute for Brain Research, Beijing 102206, China.
*Address correspondence to: Xiaorong Gao; gxr-dea@tsinghua.edu.cn and Yijun Wang; wangyj@semi.ac.cn

# Abstract
Brain-computer interface (BCI) technology establishes a direct communication pathway between the brain and external devices. Current visual BCI systems suffer from insufficient information transfer rates (ITRs) for practical use. Spatial information, a critical component of visual perception, remains underexploited in existing systems because the limited spatial resolution of recording methods hinders the capture of the rich spatiotemporal dynamics of brain signals. This study proposed a frequency-phase-space fusion encoding method, integrated with 256-channel high-density electroencephalogram (EEG) recordings, to develop high-speed BCI systems. In the classical frequency-phase encoding 40-target BCI paradigm, the 256-66, 128-32, and 64-21 electrode configurations brought theoretical ITR increases of 83.66%, 79.99%, and 55.50% over the traditional 64-9 setup. In the proposed frequency-phase-space encoding 200-target BCI paradigm, these increases climbed to 195.56%, 153.08%, and 103.07%. The online BCI system achieved an average actual ITR of 472.72±15.06 bpm. This study demonstrates the essential role and immense potential of high-density EEG in decoding the spatiotemporal information of visual stimuli.

# INTRODUCTION
The visual evoked potential (VEP)-based brain-computer interface (BCI) system is a non-invasive neural interaction technology that interprets brain activity elicited by visual stimuli, enabling the control of external devices or the transmission of information [1, 2]. With its high information transfer rate (ITR), rapid responsiveness, and minimal training requirements for users, the VEP-BCI system have gained significant attention in both academic research and consumer applications [3, 4]. With the rapid advancement of VEP encoding and decoding methods, a high-speed BCI speller reported in 2018 demonstrated the ability to input one character every 0.8 seconds, achieving a world record ITR of 325.33 bpm for non-invasive BCIs [5]. Despite these achievements, the ITR of current BCI technologies remains far from matching the efficiency of traditional human-computer interaction methods, such as touchscreens and keyboards. Moreover, since 2018, progress in improving the ITR of VEP-based BCIs has nearly stagnated, underscoring the urgent need for innovations in the field [5-27].

Existing research on visual BCIs predominantly employs commercially available 64-channel electrode caps adhering to the international 10-10 system, which provide a spatial resolution of approximately 3 centimeters [5-18, 20-22, 28-33]. However, the fundamental structural and functional units of the visual cortex comprise a complete orientation column cycle, a full set of left- and right-eye ocular dominance columns, and blobs, occupying an area of approximately one square millimeter [34]. While the "spatial Nyquist frequency" of electroencephalograms (EEG) suggests that at least 128 electrodes are sufficient for adequately sampling EEG signals across the entire scalp [35, 36], the objectives of EEG analysis often extend beyond capturing high-amplitude, low-spatial-frequency signals. Many applications require detecting finer neural activity within the visual cortex. Petrov et al. demonstrated that employing a custom ultra-high-density grid electrode array to reduce the inter-electrode spacing in the parieto-occipital scalp region (from three centimeters to one centimeter) effectively improved the signal-to-noise ratio (SNR) of VEP responses [37]. Similarly, Robinson et al. modified a 128-channel electrode cap based on the Biosemi system by increasing the electrode density in the occipital and temporal regions by two to three times [38]. Their findings confirmed the significant advantage of ultra-high-density electrodes (1.4 centimeters inter-electrode spacing) over conventional commercial electrodes (2.8 centimeters inter-electrode spacing) in capturing subtle spatial patterns of visually evoked EEG activity. These spatial characteristics are crucial for improving the decoding of visual information and hold promise for enhancing the communication speed of BCI systems.

The steady-state visual evoked potential (SSVEP) refers to the stable response of the visual system elicited by periodic repetitive stimuli [31]. Among EEG-based BCI systems, SSVEP-BCI systems exhibit the highest performance [5, 12]. Encoding methods for SSVEP-BCI systems are generally categorized into two main types: single-feature encoding and hybrid-feature encoding [1]. Single-feature encoding includes methods such as frequency modulation [21], phase modulation [27], space modulation [14] and so on. These approaches utilize the time-locked and phase-locked characteristics of SSVEP signals, along with the retinotopic mapping properties of the visual system, to encode different commands in BCI systems based on the temporal and spatial features of visual stimuli. Compared to other encoding methods, current spatial encoding strategies exhibit notable limitations in the size of visual stimuli, the number of encodable targets, and overall system performance [14, 17, 19]. For instance, Chen et al. designed a 45-target (3.5º visual angle) SSVEP-BCI speller system that utilized a frequency range of 7–15.8 Hz with 0.2 Hz intervals, achieving an ITR of 105 bpm [21]. In contrast, Maye et al. developed a nine-target BCI system employing the relative positional differences between circular flickering stimuli (27º visual angle) and the fixation point, resulting in an ITR of only 40.80 bpm [19]. Hybrid-feature encoding integrates multiple features to achieve more efficient encoding of visual targets [32]. For example, Nakanishi et al. combined phase information with frequency encoding method to enhance the differentiation of SSVEP responses between adjacent frequencies. This approach enabled the development of a 40-target SSVEP-BCI system, achieving an ITR of 325.33 bpm, which set a world record in 2018 [5]. While the reported spelling speed of 0.8 seconds per character in this study approached the upper limit of human gaze control, there remained significant potential for further optimization, particularly in increasing the number of commands (40) and improving classification accuracy (89.80%).

The visual system possesses remarkably high spatial resolution, with spatial location information consistently preserved throughout the transmission of visual information [39]. A distinct retinotopic projection exists between the primary visual cortex of each cerebral hemisphere and the corresponding hemiretina of the contralateral visual field [40, 41]. As a result, variations in the relative position between visual stimuli and the focus of overt attention produce unique topographical distributions of scalp EEG signals [8, 14, 17]. However, traditional 64-channel EEG recordings exhibit notable limitations in decoding spatial information, hindering the advancement of efficient encoding strategies that seamlessly integrate spatial information with other features such as frequency and phase.

This study aimed to enhance the ITR of visual BCI systems through the proposed frequency-phase-space fusion encoding method and high-density EEG decoding strategy. On one hand, the command set was expanded by assigning multiple fixation points as independent targets to visual stimuli flickering at specific frequencies and initial phases. On the other hand, a 256-channel EEG acquisition system was employed to capture high spatial resolution EEG signals, thereby improving the classification accuracy of the BCI system. The fundamental principles of fixation point decoding were investigated within the framework of high-density EEG. Furthermore, the effect of electrode density on the performance improvement of frequency information decoding and spatial information decoding was systematically analyzed using standard commercial EEG caps with 64, 128, and 256 electrodes. The large command-set SSVEP-BCI system developed in this study achieved an online ITR of 472.72±15.06 bpm, marking a significant breakthrough in communication speed.

## MATERIALS AND METHODS
### Overview of the high-speed BCI

To explore the potential of high-density EEG in decoding spatiotemporal information, this study designed a BCI system that encoded the targets using the rhythmic variations in visual stimulus luminance and the relative position between the stimulus and the focus of attention. Periodic visual flicker is known to induce strong SSVEPs in the parieto-occipital region [7, 20, 21]. Traditional SSVEP-BCI systems have predominantly relied on signals from nine electrode positions (Pz, POz, Oz, PO3–PO6, O1, O2) within the standard 64-channel montage for target recognition [5, 7, 10, 12, 15, 20, 30-32]. In this study, a 256-channel Quik-Cap Neo Net (mean inter-electrode distance: 1.8 centimeters) configured according to the international 10-5 system was employed [29], covering 66 channels across the parieto-occipital region. As shown in Fig. 1A, down-sampling allowed for the generation of standard Quik-Cap configurations with lower electrode density, such as 128 channels (mean inter-electrode distance: 2.8 centimeters) and 64 channels (mean inter-electrode distance: 3.1 centimeters). This study systematically compared the visual information decoding performance of four electrode configurations: 256-66, 128-32, 64-21, and 64-9.

The structure diagram of the 200-target BCI system is depicted in Fig. 1B. This study proposed a frequency-phase-space fusion encoding strategy to enhance the visual resolution requirements in decoding tasks, thereby fully leveraging the advantages of high spatial resolution EEG recordings. First, following the frequency-phase encoding method [32], 40 flickers (3.3º visual angle) with distinct frequencies (8–15.8 Hz, 0.2 Hz intervals) and phases (0–2π, 0.35π intervals) were arranged in a five × eight grid on the system interface. Unlike traditional high-speed BCI systems that assign a unique flicker to each target [5, 9, 21, 32], this study further introduced a spatial coding strategy. Five cross-shaped fixation points (0.23º visual angle) were placed along the horizontal and vertical midlines, as well as at the center of each flicker. This fusion encoding strategy allowed the system to multiplicatively increase the number of targets to 200, while maintaining the same interface size (displayed on a 24.5-inch LCD monitor). SSVEP exhibits strong time-locking characteristics with the periodic flicker stimuli. Its spectral features are marked by prominent peaks at the stimulus frequency and its harmonics, which serve as key features to distinguish different flickers

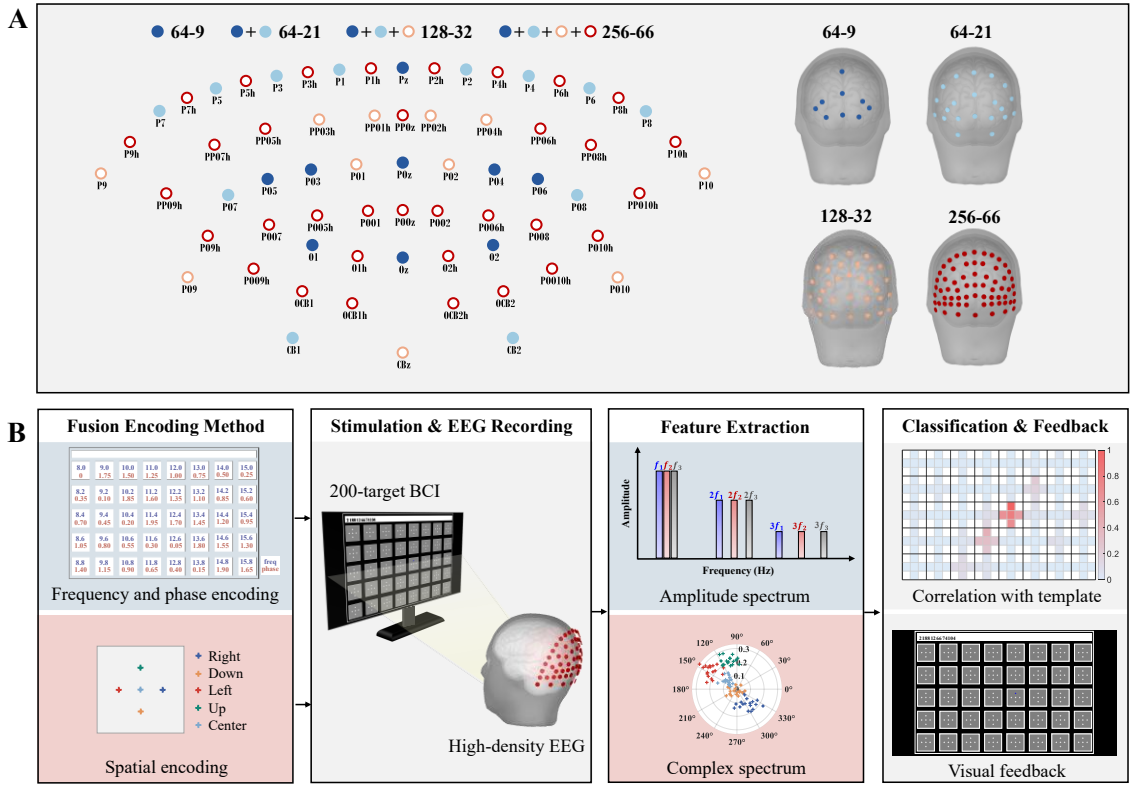

***Fig.1***. *Schematic overview of the 200-target BCI system. (A) Top and rear views of 64-9, 64-21, 128-32, and 256-66 electrode configurations. (B) Basic structure of the BCI system. The 200 targets were formed by a combination of 40 flickers with distinct frequency and phase values, along with five fixation points (right, down, left, up, center) placed within each flicker. Stimuli with high visual resolution were presented to the subjects, while high spatial resolution EEG was recorded for feature extraction. The classification results were presented to the subjects in the form of visual feedback, including color changes of selected fixation points and the display of corresponding numeric labels (Supplemental Fig. S1) in the text input box.*

[7]. A one-to-one spatial mapping exists between retinal positions and the corresponding visual cortical locations [8], enabling the SSVEP responses induced by different fixation points exhibiting distinct amplitude and phase topographic distributions. The enhancement of EEG density improved the ability to capture these spatial pattern differences. In this study, spatiotemporal filters were employed to extract task-relevant EEG components for different fixation points, whose complex spectral features exhibited a high degree of discriminability. The classification results were obtained through a template-matching approach and then provided to the participants in real-time as visual feedback.

## Experimental setup
*Participants and experimental environment*

Fifteen healthy participants (including eight females, aged 23 to 34 years) with normal or corrected-to-normal vision were enrolled in the study. Written informed consent was obtained from all participants prior to the experiment, and they were fully briefed on the procedures and requirements. The protocol conformed to the Declaration of Helsinki and was approved by the Institutional Review Board of Tsinghua University (No. 20230058). All experiments took place in a brightly lit room with stable lighting conditions. Participants were seated comfortably on a chair positioned approximately 70 centimeters directly in front of the screen. To reduce electromyographic artifacts caused by head movements, a chin rest was used to support their chin and forehead.

*High-density EEG acquisition*

EEG signals were collected from 66 electrodes in the parieto-occipital region using a 256-channel Quik-Cap Neo Net (Fig. 1A). The electrode layout adhered to the international 10-5 system, with the reference electrode positioned at the vertex and the ground electrode located at the midpoint of the FPz-Fz line. During data acquisition, all electrode impedances were maintained below 15 kΩ, and the sampling rate was set to 1000 Hz. At the onset of each visual stimulation, event triggers were generated by the stimulation program, sent to the EEG amplifier via a parallel port, and then recorded in an event channel synchronized with the EEG data. In subsequent data analysis, the 256-66 electrode configuration was down-sampled to generate electrode configurations of 32 electrodes for the 128-channel standard Quik-Cap and 21 electrodes for the 64-channel standard Quik-Cap (Fig. 1A，created using brainstorm [42]). Additionally, nine electrode positions (Pz, POz, Oz, PO3–PO6, O1, and O2) from the standard 64-channel Quik-Cap, commonly used in previous visual BCI studies [5, 7, 10, 12, 15, 20, 30-32], were also included as a control group.

*Frequency-phase-space fusion encoding method*

The frequency-phase-space fusion encoding method generated 200 targets through a combination of 40 flickers and five fixation points. The 40 flickers were arranged in a five × eight matrix, with each flicker measuring 144×144 pixels, corresponding to a visual angle of 3.3º. The horizontal and vertical spacing between adjacent flickers was 50 pixels. As illustrated in Fig. 1B, each flicker was assigned progressively increasing frequency and phase values, arranged sequentially from top to bottom and left to right. The frequency parameters ranged from 8 to 15.8 Hz with 0.2 Hz intervals, while the phase parameters varied from 0 to $2\pi$ with $0.35\pi$ intervals. Five cross-shaped fixation points (color: white) were positioned along the horizontal centerline (left and right fixation points), the vertical centerline (up and down fixation points), and the center (center fixation point) of each flicker. The distances between the right, down, left, and up fixation points and the center fixation point were uniformly 36 pixels, equivalent to one-quarter of the flicker's edge length. Each fixation point occupied a visual angle of 0.23º. The system interface was presented on a 24.5-inch liquid crystal display (Alienware 2521H) with a resolution of 1920×1080 pixels and a refresh rate of 240 Hz. The brightness values of the flickers were modulated for each frame using the sampled sinusoidal stimulation method. The stimulation program was developed with Psychophysics Toolbox under MATLAB [43].

**Experimental procedure**

This study involved offline and online experiments conducted on separate days. Fifteen subjects participated in the offline experiment, and 10 of them were called back to participate in the online experiment. During the offline experiment, all participants utilized the 200-target BCI system. Each participant completed 18 blocks of EEG data collection, with each block randomly traversing all 200 targets, corresponding to 200 trials. Each trial consisted of 0.5 seconds of cue time (one of the 200 fixation points turned red and the corresponding flicker was highlighted with a 15-pixel-wide red frame), followed by 0.5 seconds of stimulation (40 flickers flashed periodically and simultaneously). The offline data were used to determine the optimal target number and fixation combination for each participant.

The online experiment consisted of three sessions: the stimulation duration optimization session, the training session, and the testing session. During the stimulation duration optimization session, all participants uniformly used an 80-target BCI system, comprising 40 flickers and two fixation points (down and up), to streamline the optimization process. Given that the peak ITR in the offline experiment occurred at a data length of 0.2 seconds, and previous studies have shown that participants can adapt to a stimulation time of 0.3 seconds [5], the available stimulation duration for adjustment were 0.2 seconds, 0.25 seconds, and 0.3 seconds. First, six blocks of EEG data were collected under a 0.5-second cue time and a 0.2-second stimulation time. The classification performance for the 80 targets was then compared with the results obtained under the same conditions in the offline experiment. If the performance difference exceeded 10%, the stimulation duration would be extended to 0.25 seconds and another six blocks of EEG data would be collected. If a substantial performance difference still remained, the stimulation duration of 0.3 seconds would be directly adopted in subsequent experiments. During the training session, the system parameters were configured based on each participant's personalized target number, fixation combination, and stimulation duration. A total of 18 blocks of EEG data were collected, with each block randomly traversing all targets. The cue time was maintained at 0.5 seconds. During the testing session, the experimental procedure was similar to that of the training session, with the primary difference being the provision of visual feedback to participants when the online data analysis program obtained the classification results. The visual feedback consisted of color changes of the selected fixation points and the display of corresponding numeric labels in the text input box (Fig. 1B and Supplemental Fig. S1). Each participant completed five blocks of EEG data collection in this session.

**Target classification algorithm**

Task discriminative component analysis (TDCA) is a state-of-the-art algorithm widely used in SSVEP decoding tasks, which performs discriminant analysis to derive optimal spatiotemporal filters. For training data with $N_c$ electrodes, $N_p$ data points (with a time window of $[1, N_p]$), and $N_b$ training blocks, the $i$-th training trial is denoted as $X^{(i)} \in \mathbb{R}^{N_c \times N_p}$, where $i = 1, 2, \cdots N_b$. First, for each trial, the multi-channel EEG data is augmented as follows:

$$\widetilde{X} = [X^T, X_1^T, \cdots, X_l^T]^T \tag{1}$$

where $X_l$ represents the EEG data within the time window $[1+l, N_p + l]$, and $l$ represents the delay order. The augmented EEG data segment is denoted as $\widetilde{X} \in \mathbb{R}^{(l+1)N_c \times N_p}$. The same processing procedure was applied to the testing data, with the only difference being that data points exceeding $N_p$ were zero-padded. Furthermore, by incorporating prior knowledge of the stimulus frequency in SSVEPs, both the training and testing EEG data undergo secondary augmentation:

$$X_a = [\widetilde{X}, \widetilde{X}_p] \tag{2}$$

where $\widetilde{X}_p$ represents the projection of $\widetilde{X}$ onto the subspace spanned by the sine-cosine reference signals. For the $n$-th target, the template signal was obtained by averaging the training trials:

$$\overline{X}_n = \frac{1}{N_b}\sum_{i=1}^{N_b} X_a^{(i)} \tag{3}$$

where $n = 1, 2, \cdots N_t$, and $N_t$ represents the number of targets in the classification task. Subsequently, the common projection direction $W$ for all targets was determined based on the Fisher criterion:

$$\max_{W} \frac{\mathrm{tr}[W^T H_b H_b^T W]}{\mathrm{tr}[W^T H_w H_w^T W]} \tag{4}$$

where $\boldsymbol{H}_b$ and $\boldsymbol{H}_w$ denote the between-class and within-class difference matrix constructed from the secondary augmented data $\boldsymbol{X}_a$. A more detailed description of the process can be found in the paper by Liu et al [12]. To fully leverage the fundamental and harmonic components of the SSVEP response, the raw EEG signal was decomposed into multiple sub-band components using a filter bank [7]. The spatiotemporal filter $\boldsymbol{W}^{(m)}$ for the $m$-th sub-band (where $m = 1, 2, \cdots, N_m$, and $N_m$ represents the number of sub-bands) was then computed. Before calculating the correlation coefficient between the individualized template for the $n$-th target and a single testing trial, the template signal $\overline{\boldsymbol{X}}_n^{(m)}$ and testing data $\boldsymbol{X}^{(m)}$ were dimensionally reduced using spatiotemporal filters:

$$r_n^{(m)} = \rho\left(\left(\boldsymbol{X}^{(m)}\right)^T \boldsymbol{W}^{(m)}, \left(\overline{\boldsymbol{X}}_n^{(m)}\right)^T \boldsymbol{W}^{(m)}\right) \tag{5}$$

where $\rho(\boldsymbol{a}, \boldsymbol{b})$ denotes the Pearson correlation coefficient between $\boldsymbol{a}$ and $\boldsymbol{b}$. The weighted sum of the squared correlation coefficients across all sub-bands is defined as the classification feature:

$$\rho_n = \sum_{m=1}^{N_m} f(m) \cdot \left(r_n^{(m)}\right)^2 \tag{6}$$

where the weight coefficient is given by $f(m) = m^{-1.25} + 0.25$ [7]. Finally, the classification result was obtained by selecting the class with the highest $\rho_n$:

$$\tau = \arg\max_n \rho_n \tag{7}$$

**Data analysis**
*BCI performance evaluation*

EEG data segments from 66 channels were extracted based on a 140 milliseconds visual latency and subsequently down-sampled to 250 Hz. Target classification was performed using the filter bank TDCA algorithm (delay order $l$ set to four), with five sub-bands and passband frequency ranges set to [6 Hz, 90 Hz], [14 Hz, 90 Hz], [22 Hz, 90 Hz], [30 Hz, 90 Hz], and [38 Hz, 90 Hz] [7]. For the offline BCI experiment, a leave-one-out cross-validation approach was employed, where each of the 18 blocks was held out in turn as the test set while the remaining blocks served as the training set. This method was used to calculate the classification accuracy for varying target numbers (ranging from 40 to 200) and different fixation combinations (Table 1). The data lengths used for classification ranged from 0.02 to 0.1 seconds with a step size of 0.02 seconds, and from 0.1 to 0.5 seconds with a step size of 0.1 seconds. For the online BCI experiment, the TDCA spatiotemporal filter and individual templates calculated using EEG data from the training session were applied to the testing session. In addition to classification accuracy, ITR was calculated using Eq. (8), where $N$ represents the number of targets (ranging from 40 to 200), $P$ represents the classification accuracy, and $T$ represents the time required for each target selection. For the actual ITR calculation, $T$ includes both the stimulation time and the cue time; for the theoretical ITR calculation, $T$ includes only the stimulation time.

$$\text{ITR} = \left(\log_2 N + P\log_2 P + (1-P)\log_2\left(\frac{1-P}{N-1}\right)\right) \times \frac{60}{T} \tag{8}$$

*Signal feature analysis*

After averaging the 18 blocks of EEG data segments recorded during offline experiments, a fast Fourier transform (FFT) was applied to the 66 electrode positions to obtain the amplitude and phase spectra. The SNR is then calculated as the ratio of signal strength to noise strength:

$$\text{SNR} = 20\log_{10}\frac{10 \times y(f)}{\sum_{k=1}^{5}[y(f-\Delta f \times k) + y(f+\Delta f \times k)]} \tag{9}$$

*Table 1*. All combinations of fixation points under different target numbers.

| Combination label | One fixation point | Two fixation points | Three fixation points | Four fixation points | Five fixation points |
|---|---|---|---|---|---|
| 1 | right | right down | left up center | right down left up | right down left up center |
| 2 | down | right left | down up center | right down left center | |
| 3 | left | right up | down left center | right down up center | |
| 4 | up | right center | down left up | right left up center | |
| 5 | center | down left | right up center | down left up center | |
| 6 | | down up | right left center | | — |
| 7 | | down center | right left up | | |
| 8 | — | left up | right down center | — | |
| 9 | | left center | right down up | | |
| 10 | | up center | right down left | | |
| Target number | 40 | 80 | 120 | 160 | 200 |

Signal strength is defined as the amplitude value $y(f)$ at the target frequency $f$, while noise strength is defined as average amplitude value of the 10 frequencies (five frequencies on each side) adjacent to the target frequency $f$. Given that the data length was 0.5 seconds and the corresponding frequency interval $\Delta f$ was 2 Hz, the SNR was calculated only for flickers with even integer frequencies of 14 Hz. SNR and phase topographies were subsequently plotted to assess differences in response intensity and phase across various electrode positions, as well as to explore the relationship between fixation point locations and the spatial distribution characteristics of EEG responses. In addition, the 66-channel EEG data segments can be projected into one-dimensional space using the TDCA spatiotemporal filter, effectively suppressing background EEG noise while achieving data dimensionality reduction. Therefore, after applying the TDCA spatiotemporal filter to the 18 blocks of EEG data segments, FFT was performed again, and the differences in amplitude and phase distributions across different fixation points were visualized using a more intuitive method (complex spectra). Additionally, for the raw data filtered with the [6 Hz, 90 Hz] bandpass filter, the Pearson correlation coefficient between each pair of the 66 channels was calculated to assess the contribution of high-density EEG acquisition.

*Greedy search for optimal electrode combination*

A greedy search strategy was employed to obtain the optimal electrode combination and its corresponding classification performance for a given electrode number under a specific electrode configuration. Considering that most subjects achieve the maximum ITR with 80 targets, the electrode combination optimization was performed in the 80-target classification task. The data length was set to 0.2 seconds, and the initial electrode combinations were set to 64-21, 128-32, or 256-66. Step one: Initialized the electrode combination as $C$, which included all electrodes in a specific configuration. The number of electrodes was denoted as $N_c$. Step two: Discarded one electrode from the $N_c$ electrodes, generating $N_c$ new electrode combinations, and updated the number of electrodes to $N_c$ minus 1. For each electrode combination, the average classification accuracy across 15 subjects in the 80-target classification task was calculated. Updated the electrode combination $C$ to the one with the highest average classification accuracy. Step three: If the number of electrodes was greater than two, Step two would be repeated; otherwise, the search process would be terminated. Step four: Output the optimal electrode combination and its corresponding classification performance for each electrode number.

*Dynamic window classification based on TDCA*
Reducing the stimulus duration for trials with high signal quality minimizes the time spent on target selection, while extending the stimulus duration for trials with poor signal quality improves classification accuracy. Accordingly, a dynamic window classification strategy was employed to determine the optimal output time window for achieving higher ITRs. As the time window length increases, the probability of correctly classifying the target in a single trial also rises. Based on this observation, a confidence factor $c(k) = (k/50)^2$ was introduced to adjust the output probability across different time window lengths, where $k = 1, 2, \cdots, N_k$, and $N_k$ represents the number of time windows. The confidence factor was applied to the weighted correlation coefficients between the individualized template for the $n$-th target and a single testing trial:

$$\rho_{tn} = c(k) * \sum_{m=1}^{N_m} f(m) \cdot \left(r_n^{(m)}\right)^2 \tag{10}$$

The classification result was generated only when the risk cost of the output fell below a predefined threshold, defined as:

$$T(s) > -(\rho_{max} - \rho_{2nmax}) \tag{11}$$

where $\rho_{max}$ and $\rho_{2nmax}$ represent the largest and second-largest correlation coefficients among all targets, respectively. In this study, the dynamic window classification algorithm was validated on offline experimental data. The threshold $T(s)$ was defined as $T(s) = -(s \times 10^{-5})/2$, where $s = 1, 2, \cdots, N_s$, and $N_s$ represents the number of threshold values. The time window range was set between 0.1 and 0.5 seconds with an interval of 0.1 seconds, resulting in $N_k$ equal to 5. The threshold range was set from $-0.5 \times 10^{-5}$ to $-2.5 \times 10^{-4}$, with $N_s$ equal to 50. Specifically, if the data length extended to 0.5 seconds without satisfying the threshold, the classification result based on the 0.5-second data length would be adopted.

**Statistical analysis**
This study used repeated measures analysis of variance (RMANOVA) to assess whether the differences across various experimental conditions were statistically significant. Prior to analysis, the normality and sphericity assumptions of the data were checked. If the sphericity assumption was violated, the Greenhouse-Geisser correction was applied. Specifically, one-way RMANOVA was used to compare the classification performance across different fixation combinations. When statistically significant differences were observed, post-hoc pairwise comparisons were conducted with Bonferroni correction to control for Type I errors. Two-Way RMANOVA was used to examine the effects of electrode density and data length on classification performance, including main effects and interaction effects. A significant interaction effect indicated that the effect of one factor varied depending on the level of the other factor; in such cases, a simple main effects analysis was conducted using one-way RMANOVA to further explore the differences at various levels. The *p*-values reported in this study were adjusted using Bonferroni correction.

## RESULTS
### High-density EEG significantly enhances the performance of BCI
*Offline BCI performance*

The number of fixation points placed on each flicker can be varied from one to five, enabling the implementation of BCI systems with 40 to 200 targets. Table 1 presents all possible fixation combinations for different target numbers. An offline BCI experiment was conducted to determine the optimal number and placement of fixation points. A total of 15 participants took part in the experiment. The filter-bank task TDCA algorithm, combined with leave-one-out cross-validation, was applied for target recognition. Under the electrode configuration of 256-66 and data length of 0.2 seconds, the classification tasks for 40, 80, 120, 160, and 200 targets achieved the highest accuracy of 97.58% (combination 4: up), 92.59% (combination 6: down and up), 86.38% (combination 7: right, left, and up), 82.26% (combination 1: right, down, left, and up), and 73.40% (combination 1: right, down, left, up and center), respectively (Fig. 2A and Supplemental Fig. S2). For the 40-target BCI system without space encoding, the one-way RMANOVA showed no significant performance differences across the five fixation points ($p>0.05$). For the 80-target, 120-target, and 160-target BCI systems, which considered two to four fixation positions, significant performance differences were observed across different fixation combinations (Supplemental Fig. S3A).

Based on the optimal fixation combinations, the classification accuracy under different electrode densities was further analyzed (Fig. 2B). As assessed by two-way (density × data length) RMANOVA, the interaction between electrode density and data length was statistically significant for all target numbers (40-target: $F(2.10,29.35)=70.02$, $p<0.001$; 80-target: $F(2.39,35.87)=106.01$, $p<0.001$; 120-target: $F(2.39,33.52)=78.35$, $p<0.001$; 160-target: $F(2.47,34.64)=51.18$, $p<0.001$; 200-target: $F(2.06,28.84)=20.42$, $p<0.001$). In the frequency-phase encoding BCI system with 40 targets, paired t-tests with Bonferroni correction showed that the 256-66 electrode configuration significantly outperformed the other three configurations when the data length was less than or equal to 0.2 seconds ($p<0.05$, Supplemental Fig. S3B). For data lengths exceeded 0.2 seconds, all electrode configurations achieved classification accuracies above 95%, with no significant differences observed ($p>0.05$). In the frequency-phase-space encoding BCI system with 80 to 200 targets, increasing electrode density consistently improved classification accuracies across all data lengths ($p<0.05$). For example, in the 200-target BCI system, the classification accuracies for the 64-9, 64-21, 128-32, and 256-66 electrode configurations were 29.71±2.65%, 43.13±2.71%, 48.54±2.68%, and 53.82±2.87% at 0.1 seconds, and 70.81±2.57%, 81.37±1.80%, 84.27±1.71%, and 87.04±1.57% at 0.5 seconds, respectively.

Subsequently, the impact of electrode density on the actual ITR of the BCI system was evaluated. The actual ITR (measured in bits per minute, bpm) was calculated by considering the total time required for each target selection, which includes both stimulation time and gaze shift time, providing a more accurate reflection of the system's true command transmission speed [6]. With a high-density electrode configuration of 256-66, the BCI system achieved peak actual ITR values of 432.27±5.51 bpm (0.2 seconds), 470.64±8.97 bpm (0.2 seconds), 465.06±15.60 bpm (0.2 seconds), 461.69±18.19 bpm (0.2 seconds), and 416.22±14.18 bpm (0.3 seconds) for target numbers of 40, 80, 120, 160, and 200, respectively (Fig. 2C). In comparison, the commonly used 64-9 electrode configuration in previous studies achieved peak actual ITR values of 372.53±8.57 bpm (0.3 seconds),

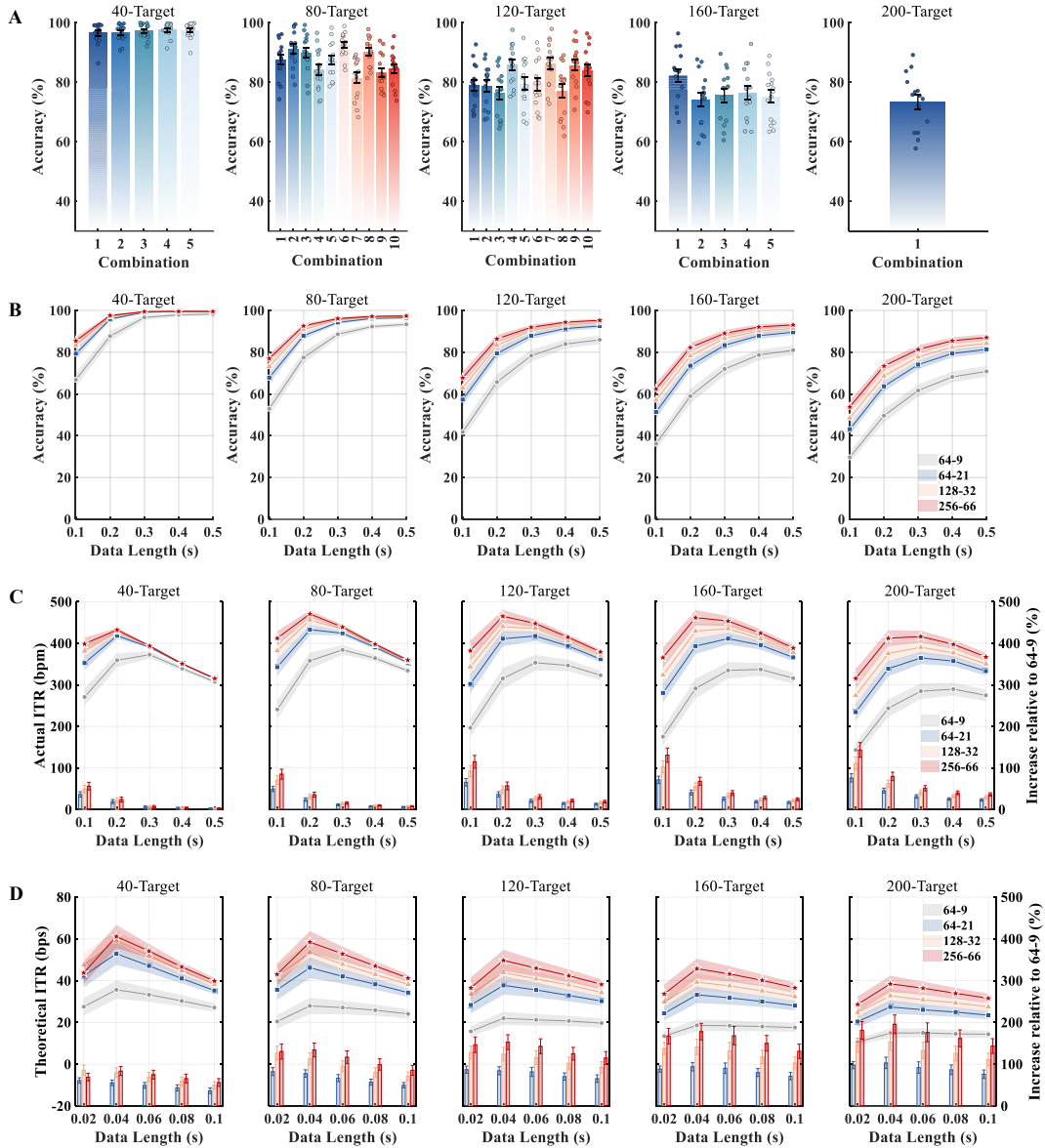

***Fig.2***. *Increasing electrode density significantly enhances visual information decoding performance. (A) Classification accuracy for all possible fixation combinations using electrode configuration of 256-66 and data length of 0.2 seconds. Error bars indicate standard errors, while discrete circles denote the individual results of 15 subjects. (B) Classification accuracy of optimal fixation combinations for four electrode configurations with data lengths ranging from 0.1 to 0.5 seconds. (C) Actual ITR (lines) for four electrode configurations, along with the ITR increase relative to the 64-9 electrode configuration (bars) for the 64-21, 128-32, and 256-66 configurations with data lengths ranging from 0.1 to 0.5 seconds. (D) Theoretical ITR (lines) for four electrode configurations, along with the ITR increase relative to the 64-9 configuration (bars) with data lengths ranging from 0.02 to 0.1 seconds. The relative increase is defined as the ratio of the difference in ITR between a given configuration and the 64-9 configuration to the ITR of the 64-9 configuration. The shaded area indicates the standard errors.*

383.98±11.56 bpm (0.3 seconds), 353.20±18.99 bpm (0.3 seconds), 337.12±15.61 bpm (0.4 seconds), and 289.64±15.59 bpm (0.4 seconds) for the 40, 80, 120, 160, and 200-target conditions. The differences in actual ITR across varying electrode densities were similar to those observed in classification accuracy. As the number of targets increased from 40 to 200, the actual ITR increase relative to the 64-9 configuration widened from 23.96%, 35.78%, 56.93%, 68.04% to 79.68% for the 256-66 configuration ($p<0.001$); widened from 22.37%, 31.28%, 48.05%, 55.58% to 62.67% for the 128-32 configuration ($p<0.001$); widened from 19.32%, 23.85%, 36.52%, 40.96% to 45.50% for the 64-21 configuration ($p<0.001$, Supplemental Fig. S3C, data length: 0.2 seconds).

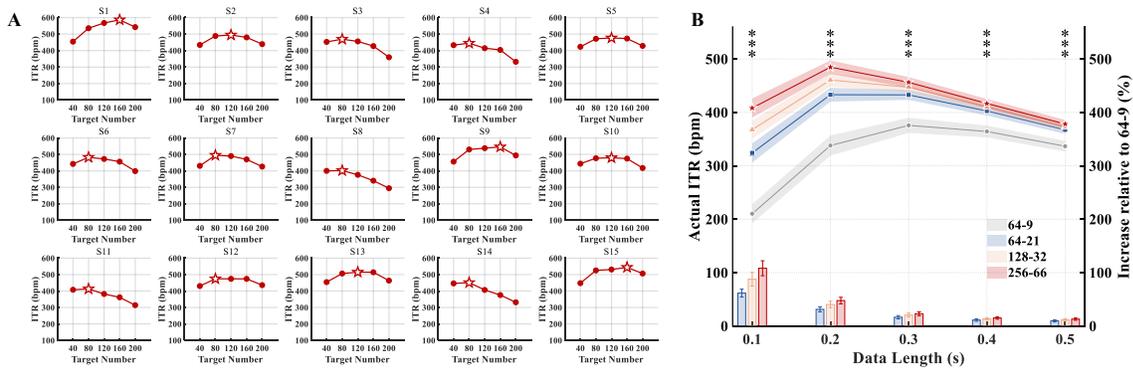

*Fig.3. Personalizing the BCI system parameters for each subject further enhances the actual ITR. (A) The relationship between the actual ITR and target number for each subject, with hollow star symbols representing the peak values. (B) The average actual ITR (lines) after personalizing the system parameters for 15 subjects, along with the ITR increase relative to the 64-9 configuration (bars). The asterisks indicate the significance level of the four electrode configurations, calculated using the one-way RMANOVA (three asterisks: p<0.001). The shaded area represents the standard errors.*

When the visual information transfer pathway was considered as a channel from a communication systems perspective, no stimulation was received during gaze shifts. Therefore, the theoretical ITR (measured in bits per second, bps) was further calculated by considering only the stimulation time under ultra-short window lengths, thereby representing the theoretical communication limits of the current BCI systems [6]. With a data length of 0.04 seconds and an electrode configuration of 256-66, the BCI systems achieved peak theoretical ITR values of 61.13±5.56 bps, 58.45±5.31 bps, 49.69±5.08 bps, 45.71±4.80 bps, and 38.45±3.92 bps for target numbers of 40, 80, 120, 160, and 200, respectively (Fig. 2D). As the number of targets increased from 40 to 200, the theoretical ITR increase relative to the 64-9 configuration widened from 83.66%, 134.02%, 152.66%, 177.95% to 195.56% for the 256-66 configuration ($p<0.001$); widened from 79.99%, 112.87%, 124.02%, 141.24% to 153.08% for the 128-32 configuration ($p<0.001$); widened from 55.50%, 77.68%, 84.27%, 93.97% to 103.07% for the 64-21 configuration ($p<0.001$, Supplemental Fig. S3D, data length: 0.04 seconds).

The average results across subjects indicated that the actual ITR increased as the target number rose from 40 to 80, but then declined consistently as the target number further increased from 80 to 200. However, the relationship between target number and actual ITR varied across individual subjects (Fig. 3A). To maximize system performance, the optimal target number and fixation combination for each subject were selected based on the highest actual ITR value (Supplemental Table S1). A total of 0, 8, 4, 3, and 0 subjects achieved peak ITR at target numbers of 40, 80, 120, 160, and 200, respectively. Among the 15 subjects, the highest actual ITR was 586.10 bpm with 160 targets, while the lowest actual ITR was 401.78 bpm with 80 targets. Fig. 3B shows the average actual ITR across 15 subjects after personalizing the system parameters. When the data length was 0.2 seconds, the actual ITR for the 256-66 configuration reached a peak of 484.76±12.68 bpm, which was significantly higher than that of the 128-32 (0.2 seconds, 460.75±12.32 bpm, $p<0.001$), 64-21 (0.2 seconds, 433.14±12.93 bpm, $p<0.001$), and 64-9 (0.3 seconds, 375.78±14.15 bpm, $p<0.001$, Supplemental Fig. S4) configurations.

*Table 2. System parameters and the corresponding performance achieved in the online BCI experiment for each subject.*

| Subject | Target number | Fixation combination | Stimulation time | Accuracy (%) | ITR (bps) | ITR (bpm) |
|---|---|---|---|---|---|---|
| S1 | 160 | right down left up | 0.25 | 96.88 | 27.57 | 551.42 |
| S3 | 80 | right left | 0.2 | 93.75 | 27.95 | 479.20 |
| S4 | 80 | down up | 0.25 | 91.00 | 21.27 | 425.45 |
| S5 | 120 | right left up | 0.2 | 88.67 | 28.08 | 481.36 |
| S7 | 80 | down up | 0.2 | 95.25 | 28.73 | 492.58 |
| S10 | 120 | right left up | 0.2 | 82.67 | 25.23 | 432.58 |
| S11 | 80 | down up | 0.2 | 87.50 | 24.95 | 427.75 |
| S12 | 80 | right left | 0.2 | 91.75 | 26.95 | 462.08 |
| S14 | 80 | down up | 0.3 | 94.75 | 18.98 | 427.05 |
| S15 | 160 | right down left up | 0.25 | 96.50 | 27.39 | 547.77 |
| Mean | — | — | — | — | 25.71 | 472.72 |
| STE | — | — | — | — | 1.02 | 15.06 |

*Online BCI performance*

To further validate the performance of the high-density EEG-based BCI system, an online experiment was conducted. Once the subjects completed a target selection, real-time classification results were provided to them in the form of visual feedback. It was important to note that, although the actual ITR in the offline experiment peaked when the data length was 0.2 seconds, not all subjects were able to switch their visual attention quickly enough to meet this standard. Therefore, in addition to the target number and fixation combinations optimized in the offline experiment, the online experiment further selected an appropriate stimulus duration for each subject. The inter-stimulus interval was uniformly set to 0.5 seconds for all subjects to cue the location of the next target. The online experiment successfully recalled 10 subjects who had previously participated in the offline experiment. Table 2 lists the system parameters and performance for each subject. The average online ITR reached 472.72±15.06 bpm.

**Principles of spatial information decoding**

The retinal-to-visual cortex mapping suggests that when visual input appears at different locations within the visual field, different regions of the visual cortex are activated [40, 41]. To characterize the spatial response differences induced by different fixation points using high-density EEG, topographic maps of response intensity and phase were depicted. Fig. 4A shows the distribution of mean SNR at the fundamental and harmonic frequencies across 15 subjects. Overall, when the fixation point was on the right, stimuli presented in the left visual field elicited a clear rightward lateralization of the response distribution at the harmonic frequencies. Conversely, when the fixation point was on the left, the lateralization was observed in the opposite direction. When the fixation point was located at the upper or lower visual field, more pronounced differences in response distribution were observed at the fundamental frequency. Specifically, with a down fixation point, the response distribution was broader and stronger near the CBz electrode, while with an up fixation point, the response distribution was more concentrated and stronger near the Pz electrode. The response phase topography for a representative subject (S1) is shown in Fig. 4B. The phase distribution at both the fundamental and harmonic frequencies also exhibited specificity based on the fixation point location. To more intuitively describe the aforementioned differences for various fixation points, the TDCA spatiotemporal filter was applied for dimensionality reduction of high-density EEG activity [12], and the low-dimensional features were subsequently visualized in the form of complex spectrum (Fig. 4C). The number and combination of fixation points were personalized based on each individual's actual ITR, with clear separation observed among different fixation points.

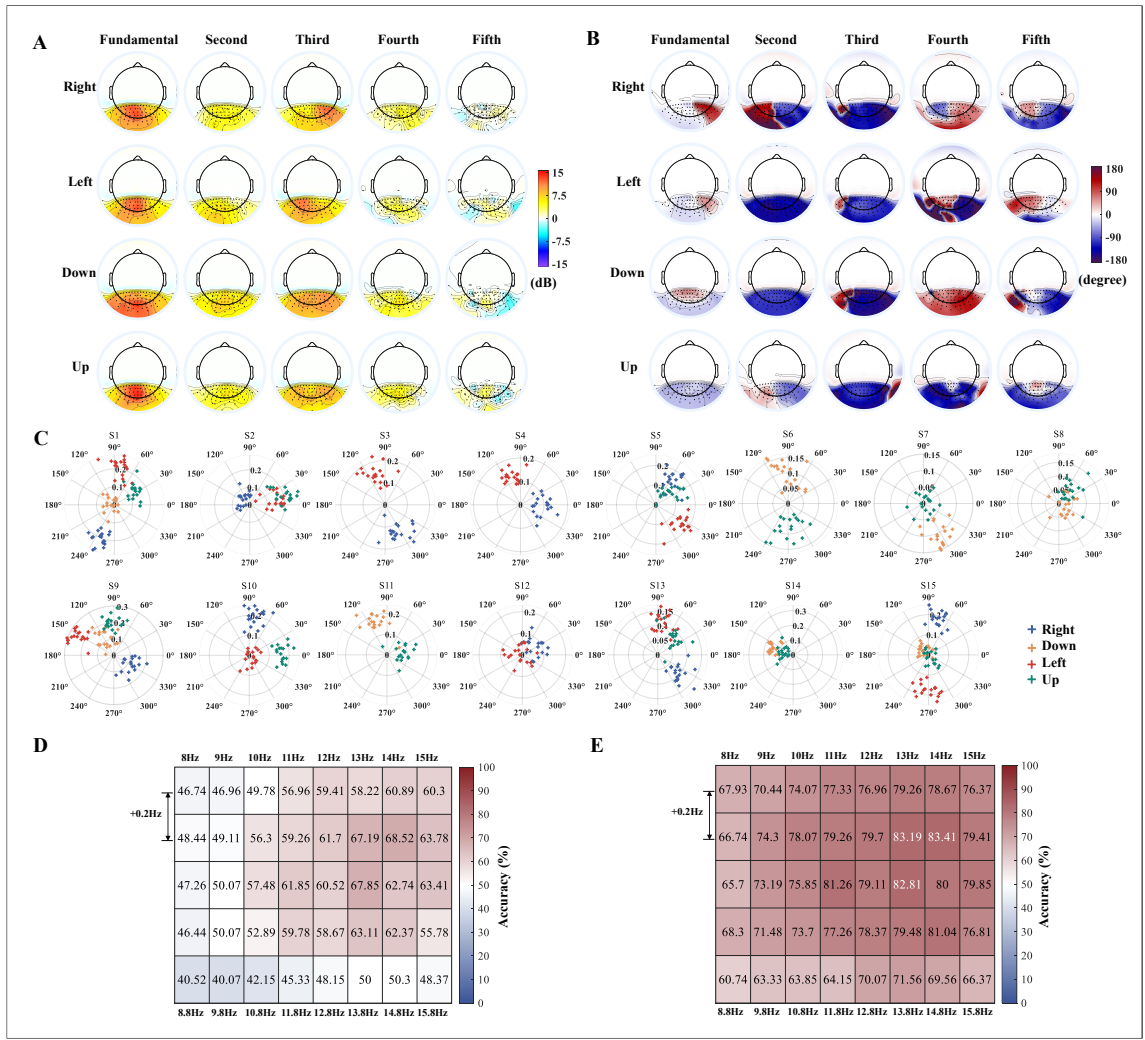

*Fig.4. Spatial response patterns and spatial information decoding performance. (A) The average SNR topographies at the fundamental and harmonic frequencies across 15 subjects for the 14 Hz flicker stimulus. (B) The phase topographies at the fundamental and harmonic frequencies for one representative subject in response to the 14 Hz flicker stimulus. (C) Complex spectrum features after TDCA spatiotemporal filtering, with each subject presenting only the fixation points selected through personalized customization. Each cross represents an individual trial, with different colors corresponding to different fixation points. (D) Classification accuracy for five fixation points on each of the 40 flickers using the spatiotemporal filter trained on the five-target classification task (data length: 0.2 seconds). (E) Classification accuracy for five fixation points on each of the 40 flickers using the spatiotemporal filter trained on the 200-target classification task (data length: 0.2 seconds).*

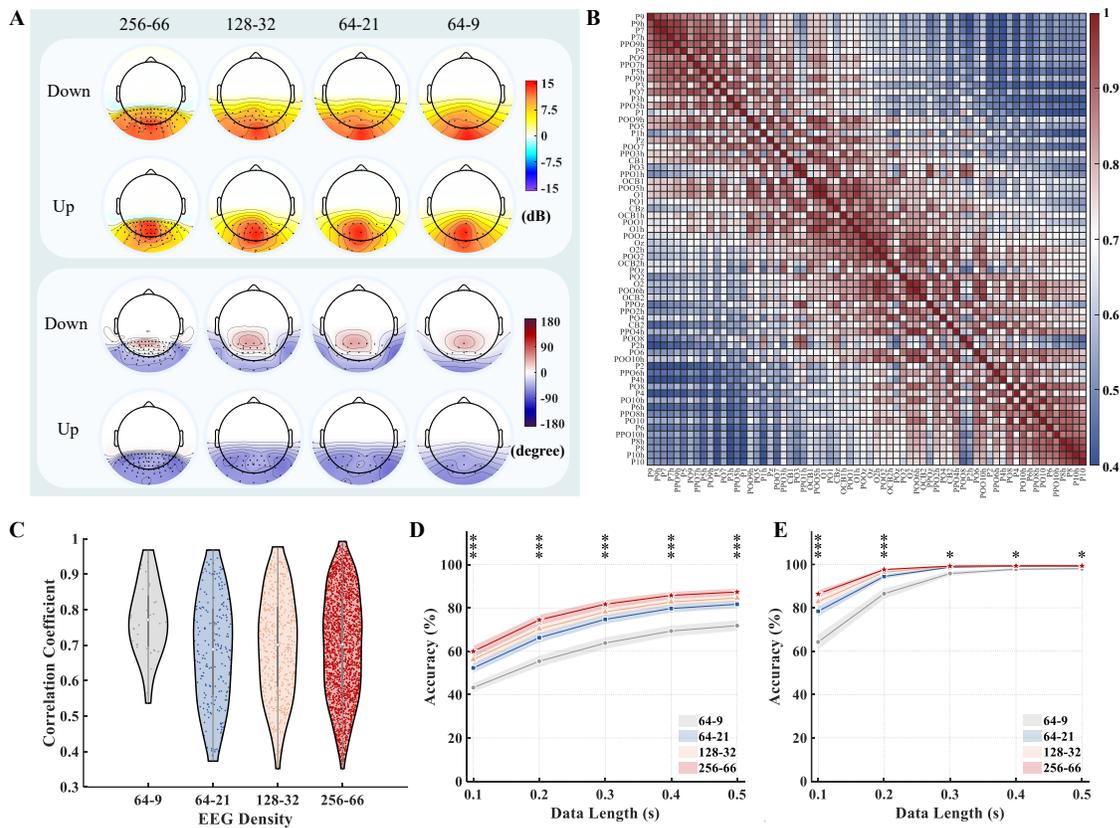

*Fig.5. High-density EEG enables finer capture of response features and more precise decoding of visual information. (A) The SNR and phase topographies at the fundamental frequency in response to the 14 Hz flicker stimulus under the 64-9, 64-21, 128-32, and 256-66 configurations. (B) Pearson correlation coefficients between all pairs of 66 electrodes under the 256-66 configurations. (C) A violin plot of the distribution of correlation coefficients under the 64-9, 64-21, 128-32, and 256-66 electrode configurations. (D) Accuracy for 5-fixation-point classification using the spatiotemporal filter trained on the 200-target classification task. (E) Accuracy for 40-flicker classification using the spatiotemporal filter trained on the 200-target classification task. The asterisks indicate the significance level of the four electrode configurations, calculated using the one-way RMANOVA (one asterisk: p<0.05, two asterisks: p<0.01, three asterisks: p<0.001). The shaded area represents the standard errors.*

Subsequently, the spatial information decoding ability under high-density EEG was further quantified. A five-fixation-point classification task was performed for each flicker. With a data length of 0.2 seconds, the average classification accuracy across 40 flickers was 54.97±1.23%. Fig. 4D shows that flickers located at the center of the interface exhibited higher classification accuracy than those at the edges, which may be due to the former's additional contribution from peripheral flickers. To further evaluate the spatial information decoding capability in the 200-target classification task, the TDCA spatiotemporal filter trained on the 200-class task was applied to the five-class task for each flicker. The average classification accuracy across 40 flickers increased to 74.47±0.98% (Fig. 4E).

**High-density EEG provides richer neural activity information**
To provide a more explicit comparison of the impact of electrode density on spatial response patterns, response intensity and phase topographies for four electrode configurations are depicted in Fig. 5A. As electrode density decreases, the contour lines in the topographies gradually smooth out, and the detailed variations in response intensity and phase in the parieto-occipital region become increasingly less discernible. Fig. 5B displays the correlation coefficients between all pairs of 66 electrodes in the parieto-occipital region. The correlation decreased as the inter-electrode distance increased. Fig. 5C shows the distribution of correlation coefficients for the four electrode configurations. Due to its smaller coverage area compared to the other three configurations (64-21: 0.68, 128-32: 0.70, 256-66: 0.70), the 64-9 configuration exhibited the highest mean correlation coefficient of 0.77. The number of electrode pairs with correlation coefficients lower than 0.6 for the 64-9, 64-21, 128-32, and 256-66 configurations were 2, 68, 140, and 610, respectively. To compare the impact of electrode density on spatial information decoding and frequency information decoding, the same spatiotemporal filter (trained on the 200-class task) was applied to process the SSVEP responses of 5 fixation points and 40 flickers, followed by classification (Figs. 5D and 5E). As assessed by two-way (density × data length) RMANOVA, the interaction between electrode density and data length was statistically significant for both the 5-fixation-point classification (F(1.52,21.25)=3.89, $p<0.05$) and the 40-flicker classification (F(1.47, 20.57)=82.89, $p<0.001$). The enhancement in spatial information decoding capability remained significant across all data lengths as electrode density increased (Supplemental Fig. S5). When the data length was 0.1 seconds, the 256-66 configuration outperformed the 64-9, 64-21, and 128-32 configurations by 16.73% ($p<0.001$), 7.65% ($p<0.001$), and 3.76% ($p<0.001$), respectively. When the data length was 0.5 seconds, the 256-66 configuration showed improvements of 15.53% ($p<0.001$), 5.65% ($p<0.001$), and 2.76% ($p<0.001$) over the 64-9, 64-21, and 128-32 configurations, respectively. In contrast, with regard to frequency information decoding capability, no significant differences were observed among most electrode configurations once a certain data length was reached. Specifically, when the data length was 0.1 seconds, the 256-66 configuration outperformed the 64-9, 64-21, and 128-32 configurations by 22.19% ($p<0.001$), 8.01% ($p<0.001$), and 3.58% ($p<0.001$), respectively. When the data length was 0.5 seconds, the 256-66 configuration showed improvements of 1.32% ($p<0.05$), 0.12% ($p>0.05$), and 0.04% ($p>0.05$) over the 64-9, 64-21, and 128-32 configurations, respectively. These results clearly demonstrated that high-density EEG enabled finer capture of response features and more precise decoding of visual information.

The greedy algorithm was used to search for the optimal number of electrodes and the corresponding electrode combinations. This method served to evaluate potential redundancy in the current electrode configuration and to compare the performance differences resulting from varying electrode densities at the same electrode number. Given that more than half of the 15 subjects achieved the highest actual ITR with 80 targets, optimization was conducted in the 80-target classification task. In the 256-66 configuration, the accuracy peaked at 52 electrodes, showing a significant difference compared to the performance with 66 electrodes (93.07% vs 92.59%, $p<0.01$, Fig. 6A). The corresponding optimal electrode combination is shown in Fig. 6B, with the unselected electrodes distributed in the peripheral areas. In the 128-32 and 64-21 configurations, the accuracy peaked at 30 and 20 electrodes, respectively, which almost included all electrodes from the respective configurations (Figs. 6C and 6D). No significant performance differences were observed compared to using all electrodes (128-32: 90.94% vs. 90.82%, $p>0.05$, 64-21: 87.94% vs. 87.88%, $p>0.05$). When the number of electrodes was fewer than 10, electrode density had no significant effect on the performance of the optimal electrode combinations ($p>0.05$; Supplemental Fig. S6). However, when the number of electrodes ranged from 11 to 21, the performance of the optimal electrode combinations searched under the 256-66 configuration was significantly better than that of the 64-21 configuration with the same number of electrodes ($p<0.05$). Similarly, when the number of electrodes ranged from 22 to 32, the performance of the optimal electrode combinations searched under the 256-66 configuration was significantly better than that of the 128-32 configuration with the same number of electrodes ($p<0.05$).

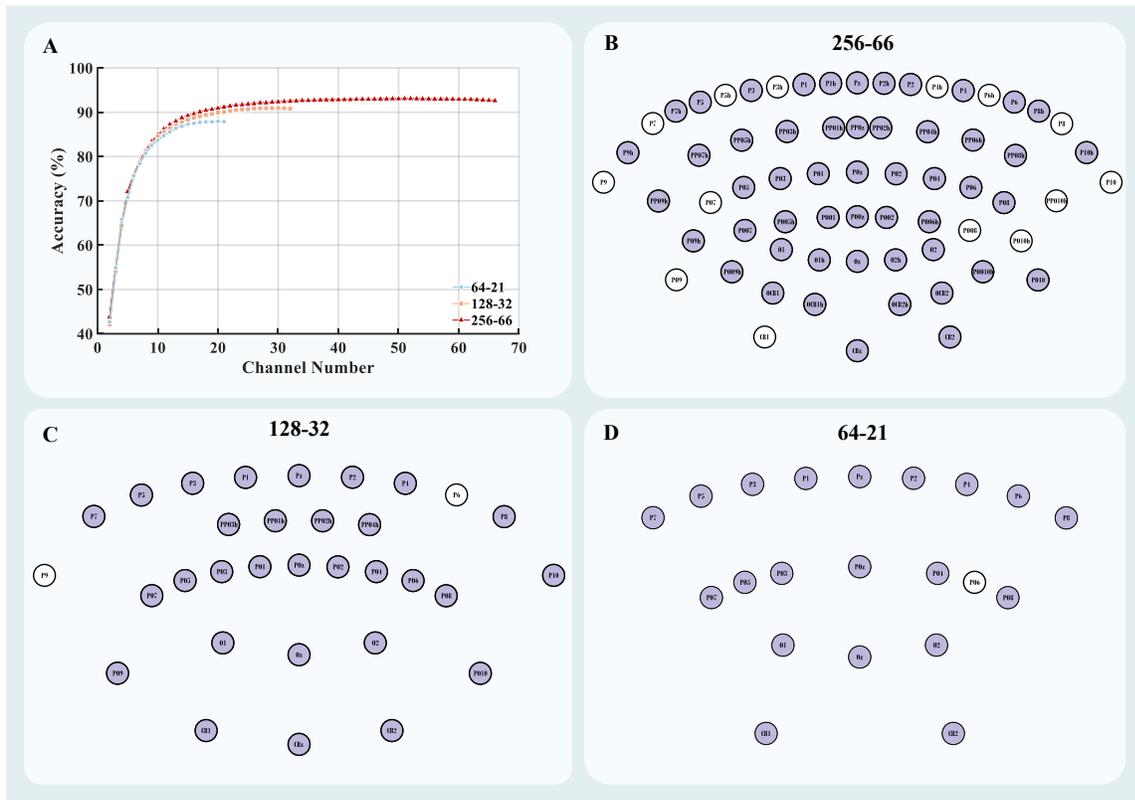

***Fig.6***. *Optimization of electrode number and combinations under different electrode densities. (A) Classification accuracy achieved by the optimal electrode combinations at different electrode number in an 80-target classification task with a data length of 0.2 seconds. (B) The optimal electrode layout for the optimal electrode number of 52 under the 256-66 configuration. (C) The optimal electrode layout for the optimal electrode number of 30 under the 128-32 configuration. (D) The optimal electrode layout for the optimal electrode number of 20 under the 64-21 configuration. Purple represents the selected electrodes, while white represents the unselected electrodes.*

**Dynamic window classification algorithm further boosts ITR**

Considering the variability in spatial information decoding capabilities across subjects, the dynamic window classification algorithm is anticipated to further enhance system performance [13]. The risk cost for outputting results was evaluated based on the difference between the first and second largest correlation coefficients of the test signal and the template signal. The output did not occur unless either the risk cost fell below the threshold or the stimulus duration reached 0.5 seconds. Therefore, the stricter the threshold, the longer the output time. Fig. 7 shows the average output time across all trials and the corresponding classification performance for different thresholds. As the number of targets increased from 40 to 200, the average output time required to achieve peak ITR using the dynamic window classification algorithm also increased (40-target: 0.151 seconds, 97.82±0.34%, 80-target: 0.168 seconds, 94.16±0.76%, 120-target: 0.192 seconds, 90.08±1.37%, 160-target: 0.216 seconds, 88.45±1.57%, 200-target: 0.250 seconds, 82.77±1.75%). Furthermore, compared to the fixed window method, the dynamic window classification algorithm demonstrated a significant improvement in peak ITR (40-target: 432.27±5.51 bpm vs. 467.28±6.69 bpm, $p <0.001$, 80-target: 470.64±8.97 bpm vs. 507.59±9.93 bpm, $p <0.001$, 120-target: 465.06±15.60 bpm vs. 502.89±15.64 bpm, $p<0.001$, 160-target: 461.69±18.19 bpm vs. 503.48±16.92 bpm, $p <0.001$, 200-target: 416.22±14.18 bpm vs. 456.72±16.87 bpm, $p<0.001$).

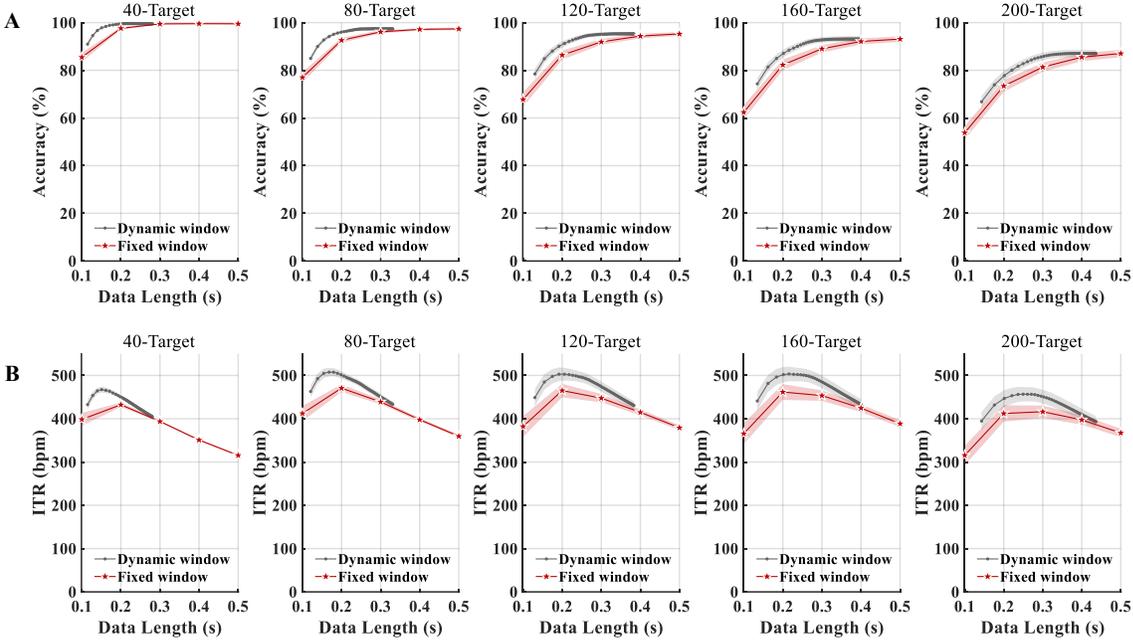

*Fig. 7.* Dynamic Window classification algorithm further enhances the performance of the BCI system. (A) Comparison of classification accuracy using fixed and dynamic window classification algorithms. (B) Comparison of actual ITR using fixed and dynamic window classification algorithms. The fixed window algorithm output classification results at five fixed data lengths, ranging from 0.1 to 0.5 seconds in 0.1-second increments (indicated by the red stars). The dynamic window algorithm determined the data length for output based on whether the risk cost fell below the threshold. The gray dots represent the average classification duration across trials at different thresholds, along with the corresponding classification performance.

## DISCUSSION

Although previous studies have confirmed that high-density EEG captures finer nuances of brain activity [28, 37, 38, 44], its potential to enhance the communication speed of visual BCI systems remains underexplored. This study innovatively proposed a frequency-phase-space fusion encoding method, aimed at investigating the advantages of high-density EEG from both frequency information decoding and spatial information decoding perspectives. Under the 256-66 high-density electrode configuration, the constructed 40, 80, 120, 160, and 200-target BCI systems achieved relatively high offline ITRs of 432.27±5.51 bpm, 470.64±8.97 bpm, 465.06±15.60 bpm, 461.69±18.19 bpm, and 416.22±14.18 bpm, respectively (Fig. 2C). After personalizing the system parameters of target number, fixation combinations and stimulus duration for individual participants, an online ITR of 472.72±15.06 bpm was achieved (Table 2), fully validating the reliability and robustness of the system's performance. The proposed SSVEP-BCI systems offer three notable advantages:

(1) Efficient integration of spatiotemporal information for large command set encoding. The number of targets that can be encoded using frequency encoding method is limited by the dominant frequency bands of the VEP response [21], while space encoding method is constrained by the number of spatial positions capable of eliciting robust responses [14, 19]. The frequency-phase-space fusion encoding method proposed in this study addressed these limitations by assigning frequency and phase parameters to multiple flickers and placing multiple fixation points within each flicker. This approach significantly increased the command set size from 40 to 200, achieving a leading number of targets among reported BCI systems [5-27].

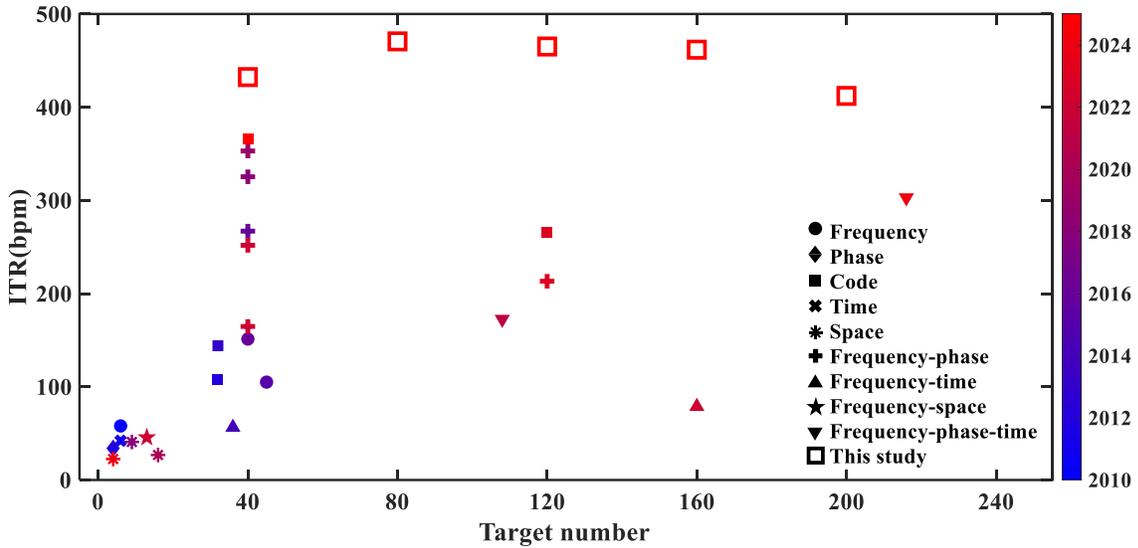

*Fig.8. Comparison of encoding methods, target number, and actual ITR of visual BCI systems in recent years with the present study.*

(2) High-resolution spatiotemporal decoding with compact visual stimuli. Traditional BCI systems typically required larger display screens to accommodate an expanded command set, which increased the difficulty for users in locating target positions [10, 11, 15]. By integrating hybrid spatiotemporal encoding method with the high-density EEG decoding strategy, this study enhanced the visual resolution of spatiotemporal information decoding. Compared to the stimulus size exceeding 3º used in the classic 40-target BCI system [5-7, 12, 33], the average stimulus size in the 80, 120, 160, and 200-target systems were reduced to 2.35º, 1.92º, 1.66º, 1.49º, thereby improving the user experience when interacting with large command set systems.

(3) Breakthrough in ITR. As illustrated in Fig. 8, the introduction of the frequency-phase encoding method and the task-related component analysis decoding strategy in 2018 marked a significant milestone in communication speed [5-27]. A 40-target SSVEP-BCI system was developed, achieving an ITR of 325.33 bpm and capable of performing a target selection every 0.8 seconds (stimulus duration: 0.3 seconds, inter-stimulus interval: 0.5 seconds) [5]. By 2024, another 40-target BCI system utilizing white noise sequence encoding method achieved the highest reported ITR of 366.05 bpm [6]. Owing to the expanded command set and the contribution of high-density EEG to feature extraction, this study achieved another breakthrough in ITR, reaching 472.72±15.06 bpm. It is worth mentioning that even the 64-9 configuration, which was commonly used in previous studies, achieved an ITR breakthrough of 383.98±11.56 bpm with 80 targets.

The optimal electrode density required to meet the demands of visual BCI systems is a key issue that has garnered significant attention in both research and application fields, yet remains challenging to answer [28, 45]. In this study, this issue required a comprehensive consideration of the encoding method, target number, and data length to draw a conclusion. In the 40-target system using frequency-phase encoding, both an increase in electrode coverage (from the 64-9 to the 64-21 configuration) and an increase in electrode density (from the 64-21 to the 256-66 configuration) resulted in significant improvements in classification performance when the data length was less than or equal to 0.2 seconds. However, when the data length reached 0.3 seconds, these improvements became negligible, and a 64-21 electrode density was sufficient. In contrast, in BCI systems with frequency-phase-space fusion encoding method, significant performance differences among the four electrode configurations were observed across all target numbers (from 80 to 200 targets) and data lengths (from 0.1 to 0.5 seconds). Furthermore, as the target number increased, the performance gap between different electrode configurations became more pronounced (Fig. 2). Given that this study focuses on basic visual information, such as periodic brightness changes and the relative position of the fixation point and visual stimuli, it can be inferred that as the difficulty of the decoding task (e.g., increased target numbers) or the complexity of the visual stimuli (e.g., natural image stimuli) increases, the demand for higher electrode density will also rise.

As the electrode density increased, subtle variations in the response characteristics across electrodes could be captured (Figs. 5A), which facilitated the enhancement of visual information decoding performance. A comprehensive comparison was made between the contributions of high-density EEG to both frequency information and spatial information decoding. Previous studies have shown that SSVEPs induced by different frequencies exhibit similar spatial distributions, concentrated in the parieto-occipital regions without any lateralization [5]. However, differences in the relative spatial positions between the fixation point and visual stimuli could lead to varying degrees and directions of lateralized amplitude distributions, as well as differentiated phase distributions in the induced SSVEPs (Fig. 4). This explained the principle underlying the distinguishability of fixation points within the same flicker under identical temporal parameters (frequency and phase values) and elucidated why increasing electrode density results in more significant improvements in spatial information decoding tasks compared to frequency information decoding (Figs. 5D and 5E).

In spatial information decoding tasks, significant performance differences existed among different fixation combinations (Fig. 2A). Interestingly, the fixation combination that yielded the best classification performance corresponded to those with the farthest physical distance. Future research could increase the number of fixation points and employ microneedle electrode arrays with millimeter-level spatial resolution [46] to precisely map spatial locations to ultra-high-density EEG response features. In addition to high-density acquisition technology, EEG source imaging methods offer another effective approach for enhancing spatial resolution by employing forward problem modeling and inverse problem solving to accurately project EEG data onto cortical voxels. Furthermore, beyond spatial locations, more complex spatial features such as color, shape, brightness contrast should also be taken into account. Developing predictive models that link visual stimulus spatial attributes with EEG responses could enable a transformative leap in spatiotemporal information encoding and decoding—from simple brightness modulation stimuli to sophisticated patterned stimuli, and ultimately to natural image stimuli. Ultimately, such advancements will drive the evolution of visual BCI technology from efficient interaction toward truly natural interaction.

## ACKNOWLEDGMENTS

**Author contributions:** Conceptualization: G.M., X. G., Y. W. Methodology: G.M., S. T., X. G., Y. W. Investigation: G.M., S. T. Visualization: G.M. Supervision: W. P., X. C., X. G., Y. W. Writing—original draft: G.M., S. T. Writing—review & editing: W. P., X. C., X. G., Y. W.

**Funding:** This work was supported in part by the National Natural Science Foundation of China under Grant 62071447, 62401325, 62201321, in part by the National Key Research and Development Program of China under Grant 2022YFF1202303, 2023YFF1203702 and in part by the Postdoctoral Fellowship Program of CPSF under Grant Number GZC20240864.

**Competing interests:** The authors declare that there is no conflict of interest regarding the publication of this article.

## DATA AVAILABILITY

The dataset for the 200-target BCI paradigm employing frequency-phase-space coding is available at https://figshare.com/s/c962d29d7752e27176be. Any additional requests for information can be directed to, and will be fulfilled by, the corresponding authors.

## SUPPLEMENTARY MATERIALS

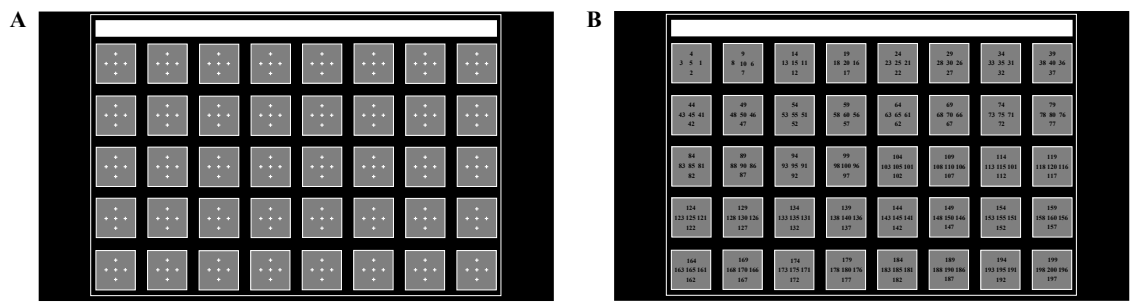

*Fig. S1*. User interface of the 200-target BCI system and corresponding numeric labels for each target. (A) The user interface consisting of a text input box and 200 fixation points. (B) The numeric label corresponding to each fixation point. In the online experiment, this label is provided as feedback and displayed in the text input box at the top of the interface.

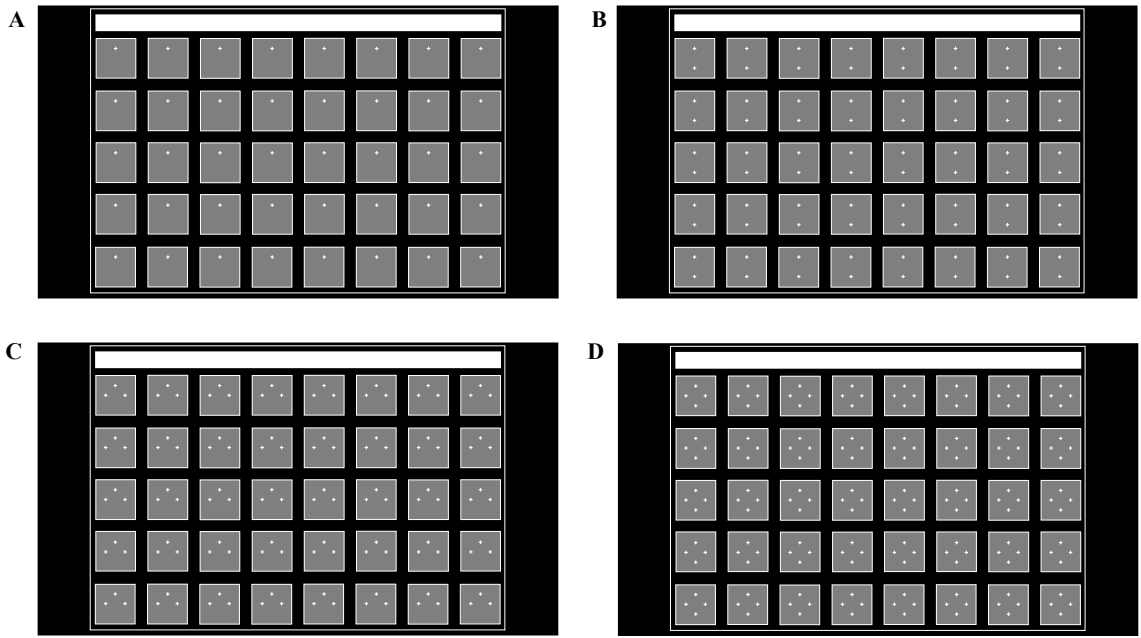

***Fig. S2***. *User interfaces of the 40, 80, 120, and 160 target BCI systems with optimal fixation combinations. (A) 40-target BCI system with fixation point of up. (B) 80-target BCI system with the fixation combination of down and up. (C) 120-target BCI system with the fixation combination of right, left, and up. (D) 160-target BCI system with the fixation combination of right, down, left, and up.*

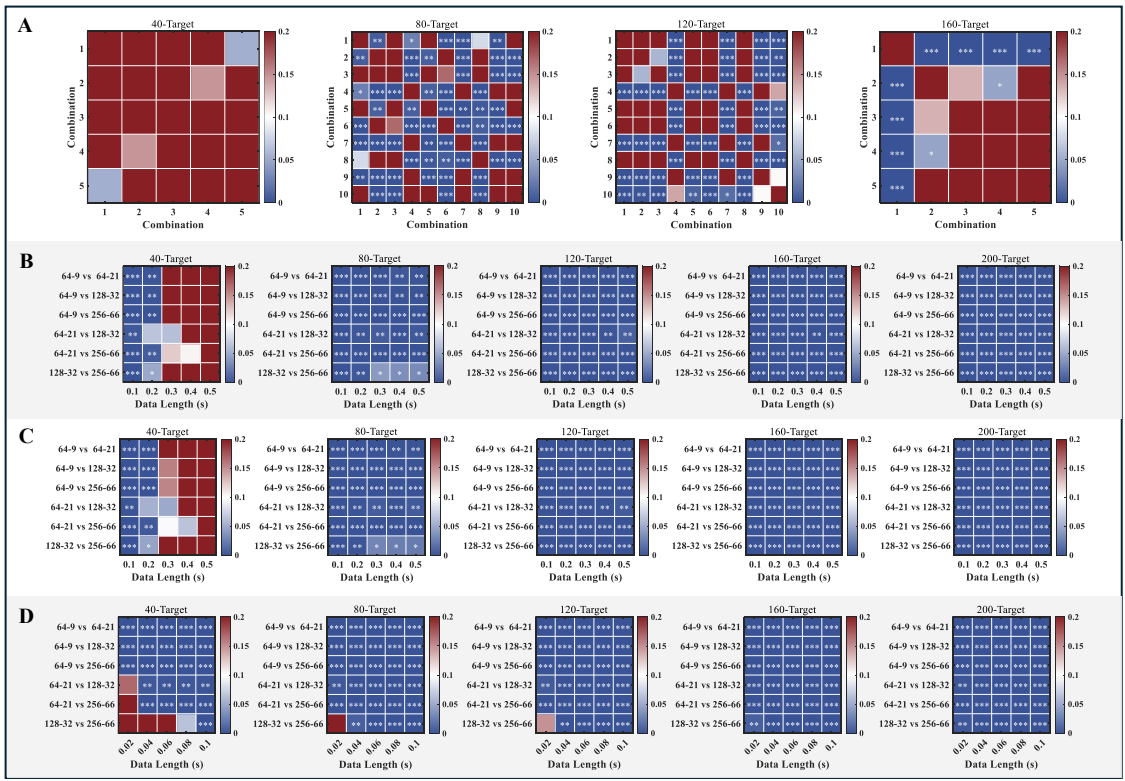

***Fig. S3***. *Paired t-test with Bonferroni correction showing significant performance differences between different electrode densities in the BCI systems. (A) Significance levels between different fixation combinations in the BCI systems with 40, 80, 120, and 160 targets. (B) Significance levels of classification accuracy for four electrode configurations with data lengths ranging from 0.1 to 0.5 seconds. (C) Significance levels of actual ITR for four*

electrode configurations with data lengths ranging from 0.1 to 0.5 seconds. (D) Significance levels of theoretical ITR for four electrode configurations with data lengths ranging from 0.02 to 0.1 seconds (one asterisk: p<0.05, two asterisks: p<0.01, three asterisks: p<0.001).

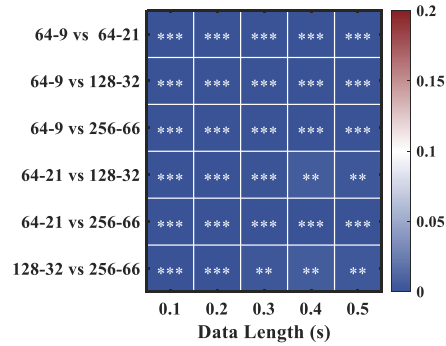

*Fig. S4*. Significance levels of actual ITR for four electrode configurations after personalizing the system parameters. A paired t-test with Bonferroni correction was performed. One, two, and three asterisks indicate $p<0.05$, $p<0.01$, and $p<0.001$, respectively.

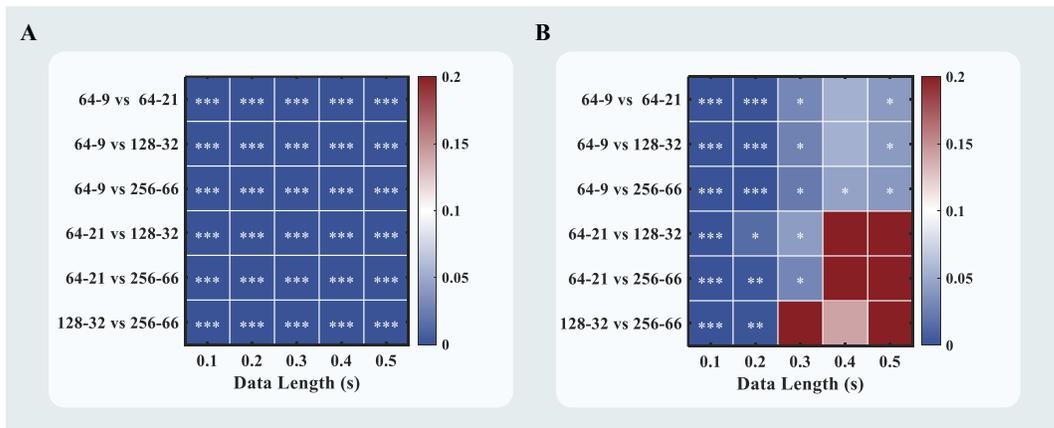

*Fig. S5*. Significance levels of accuracy for four electrode configurations under the spatial information decoding and the frequency information decoding tasks. (A) Significance levels of accuracy for four electrode configurations under the 5-fixation-point classification tasks. (B) Significance levels of accuracy for four electrode configurations under the 40-flicker classification tasks. A paired t-test with Bonferroni correction was performed. One, two, and three asterisks indicate $p<0.05$, $p<0.01$, and $p<0.001$, respectively.

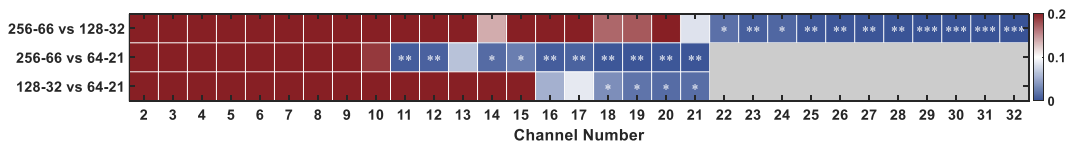

*Fig. S6*. Significance levels of accuracy for three electrode configurations under different number of electrodes. A paired t-test with Bonferroni correction was performed. One, two, and three asterisks indicate $p<0.05$, $p<0.01$, and $p<0.001$, respectively.

*Table S1. Target number and fixation combinations selected for each subject based on the principle of maximizing the actual ITR.*

| Subject | Target number | Fixation combination | Stimulation time | Accuracy (%) | ITR (bps) | ITR (bpm) |
|---|---|---|---|---|---|---|
| S1 | 160 | right down left up | 0.2 | 96.42 | 34.19 | 586.10 |
| S2 | 120 | right left up | 0.2 | 90.19 | 28.84 | 494.32 |
| S3 | 80 | right left | 0.2 | 92.57 | 27.36 | 469.01 |
| S4 | 80 | right left | 0.2 | 89.65 | 25.95 | 444.84 |
| S5 | 120 | right left up | 0.2 | 88.10 | 27.80 | 476.58 |
| S6 | 80 | down up | 0.2 | 94.10 | 28.13 | 482.25 |
| S7 | 80 | down up | 0.2 | 95.49 | 28.86 | 494.74 |
| S8 | 80 | down up | 0.2 | 84.10 | 23.44 | 401.78 |
| S9 | 160 | right down left up | 0.2 | 92.26 | 31.81 | 545.37 |
| S10 | 120 | right left up | 0.2 | 88.47 | 27.98 | 479.69 |
| S11 | 80 | down up | 0.2 | 85.56 | 24.08 | 412.77 |
| S12 | 80 | right left | 0.2 | 93.19 | 27.67 | 474.37 |
| S13 | 120 | right left up | 0.2 | 92.50 | 30.03 | 514.75 |
| S14 | 80 | down up | 0.2 | 90.42 | 26.31 | 451.04 |
| S15 | 160 | right down left up | 0.2 | 92.08 | 31.72 | 543.75 |
| Mean | — | — | — | — | 28.28 | 484.75 |
| STE | — | — | — | — | 0.74 | 12.68 |